# Heuristic algorithms in Evolutionary Computations and modular organization of biological macromolecules: applications to directed evolution

Alexander Spirov[1,2], Ekaterina Myasnikova[3]

[1]I. M. Sechenov Institute of Evolutionary Physiology and Biochemistry Russian Academy of Sciences, pr. Torez 44, St. Petersburg 194223, Russia

[2]The Institute of Scientific Information for Social Sciences RAS, Moscow, Russia

[3]Peter the Great St. Petersburg Polytechnical University, 29 Polytechnicheskaya St. Petersburg 195251, Russia

## Abstract

A while ago, the ideas of evolutionary biology inspired computer scientists to develop a thriving nowadays field of evolutionary computing, in general, and genetic algorithms (GA), in particular. At the same time, the directed evolution of biological macromolecules (in vitro evolution) is reasonably interpreted as an implementation of GA in biochemical experiments. One of the theoretical foundations of GA, justifying the effectiveness of evolutionary search, is the concept of building blocks (BB). In evolutionary computations, it is reasonable to match these BBs to domains and motifs of macromolecules in evolutionary and synthetic biology. Computer scientists have shown and carefully studied the importance of identifying and preserving already found BBs for the effectiveness of evolutionary search. For this purpose, dozens of procedures and algorithms have been developed, including crossover heuristic algorithms. On the other hand, the experimental procedures defining and preserving domains and motifs remain a poorly developed area in the techniques of evolution in vitro. In this paper, we demonstrate how several simple algorithms preserving the BBs can increase the efficiency of in vitro evolution in numerical experiments by almost an order of magnitude. As test problems, we propose and use such well-known problems of synthetic biology as the evolutionary search for strong bacterial promoters (with several conservative motifs) and search for multi-domain RNA devices, as compared to the classic GA tests (Royal Road functions). The success of these primary tests with simple and well-known algorithms gives us every reason to expect that the implementation and application of more advanced and modern evolutionary camputation procedures will give an even greater increase in efficiency. Such an increase in search efficiency will significantly reduce the cost of in vitro evolution experiments, which will fully cover the costs of developing new experimental procedures based on these algorithms.

# 1 Introduction

Since the 60s of the XX century, the transfer of ideas from biology to applied mathematics has served to form the vast and thriving field of Evolutionary Computation, EC [Holland, 1975; Koza et al., 1999]. This is an impressive example of the formation of a new field at the intersection of two sciences.

The field of Evolutionary Computation (EC) has been inspired by ideas from the classical theory of biological evolution. Genetic Algorithms (GA; a subset of EC a class of Evolutionary Algorithms) have five particular elements: encoding (the "chromosome"); a population; a method for selecting parents and making a child chromosome from the parents' chromosomes; a method for altering the child's chromosomes (mutation and crossover/recombination); criteria for fitness; and rules, based on fitness, by which offspring are included into the population (and parents retained).

In turn, impressive progress in EC field in understanding the reasons for efficiencies in evolutionary searches can be useful in life science. The last couple of decades have seen a clear reverse transfer of concepts and approaches from EC to synthetic and systems biology [Holloway, Spirov, 2012]. Namely, it has begun to influence scientific work in the field of molecular evolution and in the modeling of biological evolution [Stemmer, 1994a,b; van Nimwegen et al. 1997; 1999; Crutchfield & van Nimwegen, 2001; Voigt et al., 2002]. Moreover, the approaches of such a rapidly developing field of modern biology as the directed (forced) evolution of macromolecules can be reasonably treated as the experimental implementation of GA [Holloway, Spirov, 2012].

A number of key problems of the directed evolution of modern synthetic biology certainly belong to the field at the intersection of biology and computer science. At the same time, the area of directed evolution of biomolecules (RNA, DNA and proteins) suffers from a lack of effective mathematical and computer approaches required to reduce the cost of experimental molecular evolution and improve its efficiency [Schuster, 2006]. This makes the purposeful transfer of algorithms and approaches from EC to this biological field extremely promising.

In this paper, we will discuss how developments in the area of crossover operators for GA, provide new understanding of evolutionary search efficiencies, and the impacts this can have for biological molecular evolution, including directed evolution in the test tube. Namely, we will focus is on the substantial efficiencies that can be found in the alteration of the child chromosome by crossover.

## 1.1 Problems of directed evolution in synthetic biology

The problems of directed evolution in synthetic biology are most evident on the example of RNA-devices. RNA-devices are multi-modular RNA molecules, including sensory and effector blocks, connected by transmitter modules. They are designed so that they are able to sense a highly specific signal from environment and turn on or switch the corresponding function [ ]. RNA-devices are currently typically produced by the methods of in vitro evolution (first of all, by the SELEX approach), followed by in vivo evolution methods [Win et al., 2009].

Not only the methodology, but also the character of the selected RNA devices (their modularity) bring this experimental area closer to EC. Modularity is the basis for the design of RNA devices and it dictates approaches to their development: plug and play strategy [Spirov, Holloway, 2012, Schuster, 2006, Chappell et al, 2015].

The modularity of RNA devices has direct analogies to the concept of Building Blocks (BB) in GA and GP. In EC, we can find extensive studies on the role of BB in evolutionary search and on approaches to manipulating BB in order to improve the efficiency of evolutionary search [De Jong, 2006]. These approaches are reasonable to try to transfer and implement into these areas of synthetic biology and biotechnology.

In the field of RNA-devices synthesis the rational design approach prevails. Namely, available at hand natural modules, as well as synthetic modules (typically derived from the natural ones), are stitched together into a new molecule, usually poorly functioning or not functioning at all. Then they use the approaches of experimental evolution in vitro, followed by in vivo evolution with the goal of finally obtaining a device capable of working in living organisms. These stages is also called RNA device tuning [Liang et al., 2011].

In addition to tuning, similar approaches are used for experimental evolutionary search of devices with new functions, based on existing devices [Schuster, 2006, Chappell et al, 2015, De Jong, 2006]. This technique is called a modification of the device. The in vitro evolution succeeds, for example, in changing the specificity of the sensory domain so that it acquires sensitivity to another environmental signal.

Tuning and modification of the device are typically carried out by creating libraries of partially randomized sequences for their subsequent use as initial populations in the evolutionary search in vitro. The search can be carried out using error-prone PCR, which corresponds to a point mutation algorithm in EC. On the other hand, EC offer a whole range of procedures (primarily, recombination procedures) to significantly accelerate the evolutionary search.

Typically, RNA devices are designed by integrating existing underlying components (sensor, communication module, regulatory RNA) into a functional device framework [McKeague et al., 2016]. Alternatively, a single component in the context of the device framework may be screened to identify a sequence that functionally couples with the other components in the device. The most common design strategies focus on functionally connecting the sensor and regulatory components (e.g. through the design of an appropriate communication module) using either rational design, screening or selection strategies. However, there exists one case where both the sensing and regulatory RNA were simultaneously engineered de novo [Robertson, Ellington, 1999].

The length of the multivalent aptamer or a small aptamer biosensor will be at least several hundred bp. This corresponds to the astronomical number of possible sequences of this length. In the vastness of such a sequence space, it is probably possible to find nanodevices with practically innumerable and even unimaginable properties today. On the other hand, a rational design allows us to construct today several hundred (maximum several thousand) nanodevices based on several dozen characterized molecular modules. This, we are convinced, is the strongest argument why we need to work in the field of searching from scratch for multivalent aptamers and multimodular RNA devices.

Typically, an in vitro evolution search is understood as a search for a single functional sequence, while the design of multi-domain molecules and molecular devices is understood as a successful “stitching” of several required domains. In fact, the design of new devices typically requires an evolutionary search not only, but several modules, and preferably in one series of experiments.

Illustrative examples of the evolutionary search of several domains in parallel, this is, first of all, an accidental find of three domains in the desired regulatory element of Turnip Crinkle RNA virus [Zhang, Simon, 2003]. In these experiments, ~ 28 n.p. the viral regulatory element {motif1-hairpin} was randomized and then subjected to selection in plants (in vivo SELEX). Most of the "winners" in this experiment contained up to three short motives (~ 5-7 np), many of which are found in the naturally-known promoter elements of this virus.

Another important series of examples for us is the evolutionary search procedure for multivalent (chimeric) aptamers (and multivalent antibodies) [Burke, Willis, 1998; Wu, Curran, 1999; Ahmad et al., 2012; Vorobyeva et al., 2016].

In this approach, parallel selection for affinity for two different agents is first performed [Wu, Curran, 1999]. Then, the selected aptamers are cross-linked in a new population of chimeric molecules. Usually, in this case, chimeras lose high affinity for target molecules to which the original "halves" were selected (one aptamer interferes with the other in the "chimeric" molecule). In the next round of selection, chimeras are selected for affinity simultaneously to both targets and, as a result, bivalent aptamers are found. As Vorobyova et al. write, this so-called "chimeric" SELEX [Burke, Willis, 1998; Ahmad et al., 2012] includes the use of different combinations of aptamer domains together with the randomization of the linker [Vorobyeva et al., 2016]. After several selection rounds, the population is enriched with molecules which retain binding activities of all monomers in the context of multivalent construct.

Another important example is the parallel selection of three spacer sequences of bacterial promoters [Jensen, Hammer, 1998]. In this approach, the known conservative promoter components (elements «-44», «-35», «-15» & «-10») remained unchanged, but the spacers between them were randomized. As a result of such parallel selection of three promoter sequences, it was possible to optimize the desired "linkers".

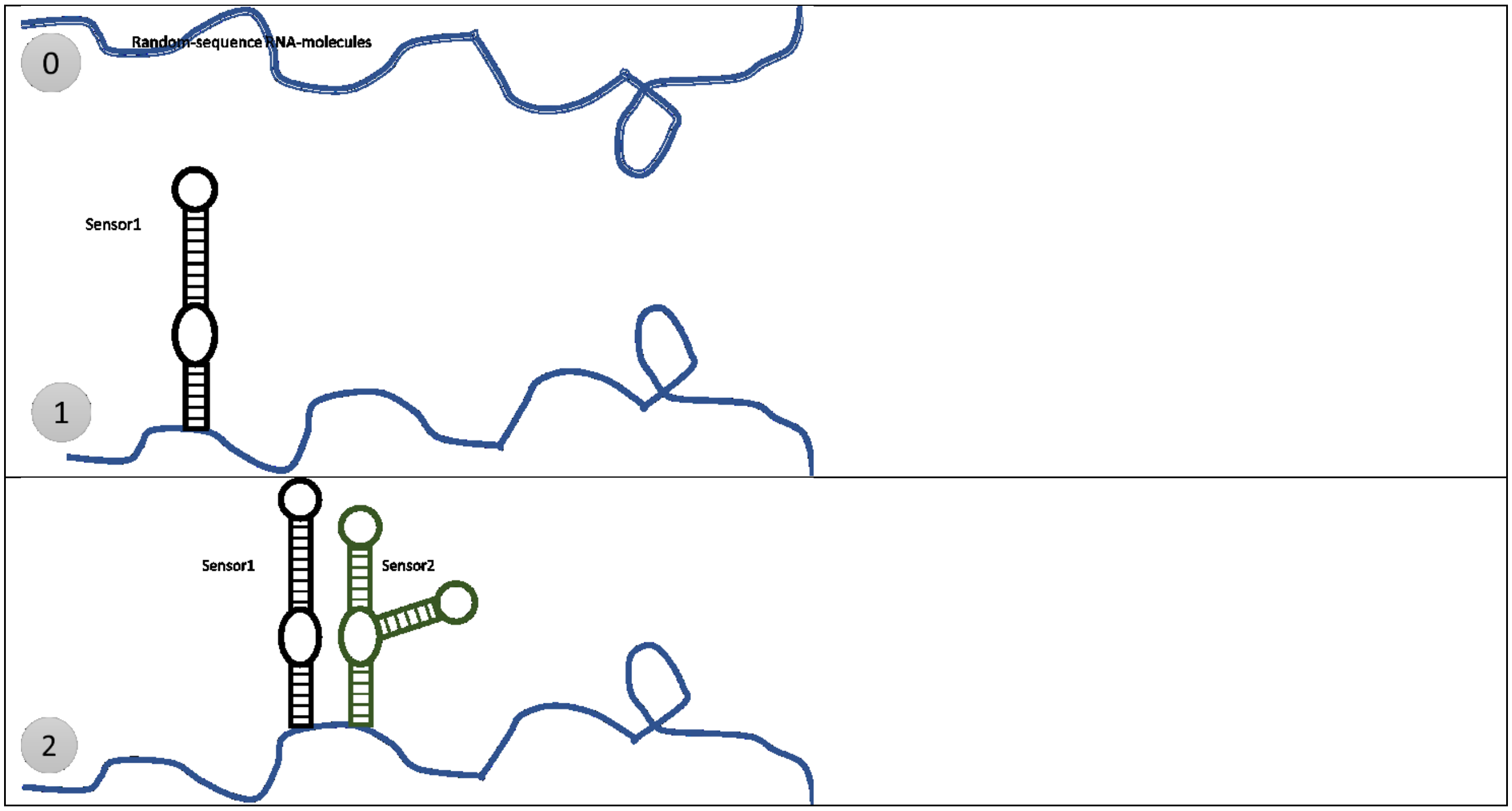

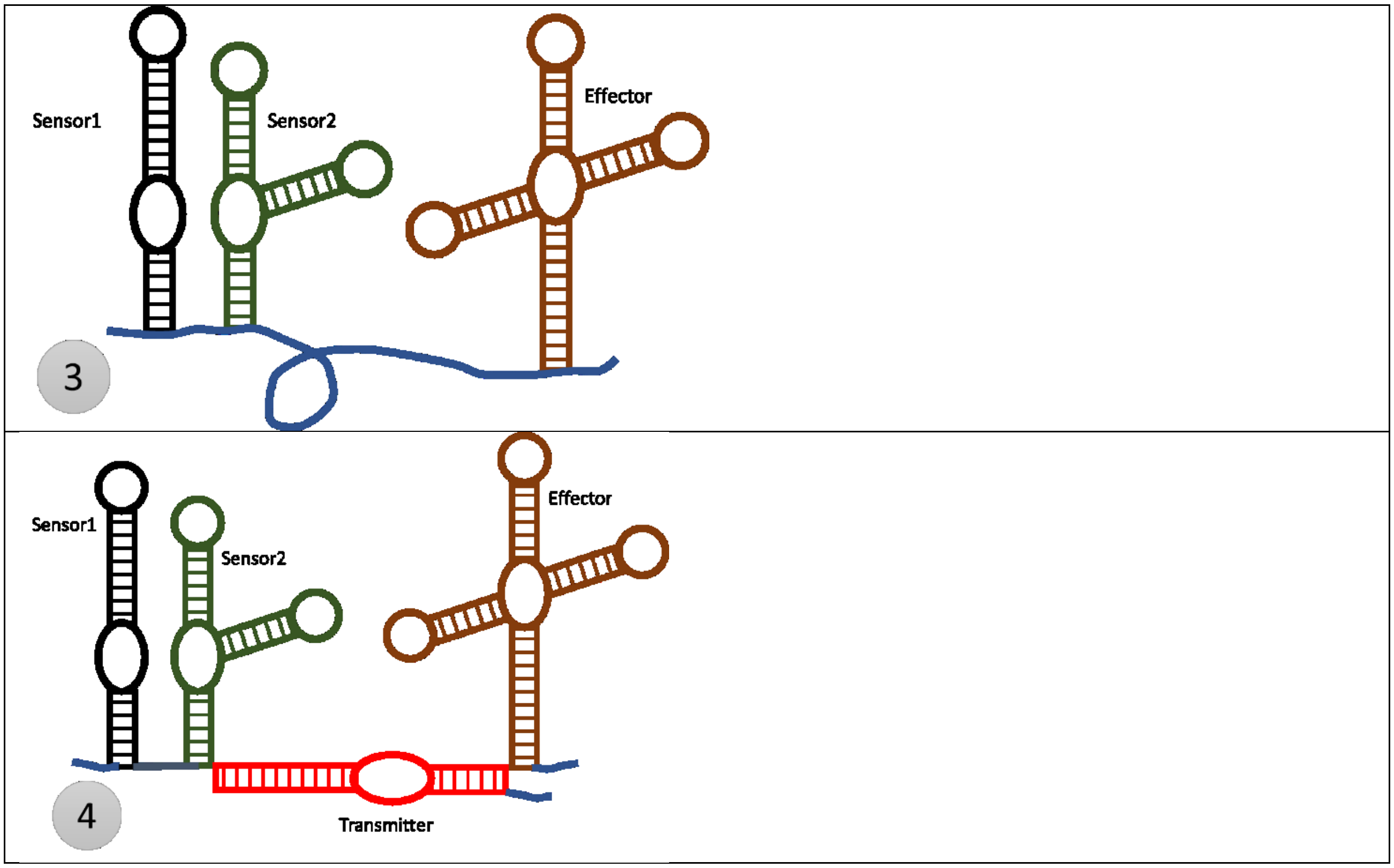

Figure 1. The general idea of sequential selection of motifs for the RNA device (from scratch). For example, sensors (or an actuator) are first selected, then an actuator (effector), and finally a transmitter (= connector).

Of course, a resource-intensive and limited analysis by real experiments on SELEX and, more broadly, directed evolution, should be preceded by an extensive numerical, computer analysis. Such an analysis should also accompany real experiments. It is precisely the prospects for the development of numerical approaches to the evolutionary design of macromolecular nanodevices that will interest us.

## 2 Recombination and modular organization of macromolecules

We use the task of modeling the evolutionary origin of a very strong bacterial promoter (using bacterial rrnP1 as an example) as an illustration that is close to real life problems. We will complete her analysis of the results of the analysis of the evolution of the problems of Royal Roads and Royal Staircases. In turn, this problem is a connecting link to tot a more realistic problem of the evolutionary design of macromolecular devices.

We formulate the problem of RNA devices in the most general form as the problem of sequentially finding the domains of the required activity (aptamers, switches, ribozymes), so that each domain is found within its own segment of the common RNA molecule (as illustrated by Fig.1). By finding all the domains we are looking for, it is supposed to optimize spacer sections (as was done in those publications on promoters and on two-domain aptamers). The last task is beyond the scope of this article.

The following general considerations allow us to make estimates by numerical experiments only on the lower boundaries of the sizes of functionally active RNA domains / motifs (they range from a minimum five-nucleotide ribozyme to 18-19 nucleotide aptamers and small functional RNAs). We will have to accept these restrictions on the minimum length of functional motives due to the almost exponential growth of the difficulty of such computational problems.

Specifically, we will choose domains in 5n (GUGGC), 11n (M4U: AAAA . . . AGUAGUC), 18n (hsa-miR-516a-3p: UGCUUCCUUUCAGAGGGU) and 19n (Inteleukin-6 eceto: GGGGAGGCUGUGGUGAGGG), separated by structural sequences from 15 to 5 n. (We keep in mind that aptamers are small and usually from 20 to 60 nucleotides [Lakhin et al., 2013]).

## 2.3 Retroviral recombination and molecular breeding

DNA shuffling is the first reported method of in vitro recombination of homologous genes invented by Willem Stemmer [Stemmer, 1994a,b]. The genes to be recombined (parental genes) are randomly fragmented to collect fragments of the desired size. These fragments are then reassembled using cycles of denaturation, annealing, and extension by a polymerase (DNA fragments with sufficient overlapping homologous sequence will anneal to each other and are then extended by DNA polymerase). Recombination occurs when fragments from different parents anneal at a region of high sequence identity. Several rounds of this PCR extension are allowed to occur, after some of the DNA molecules reach the size of the parental genes. Following this reassembly reaction, PCR amplification with primers is used to generate full-length chimeric genes.

DNA shuffling (or ‘sexual PCR’) is well advanced field now, and have been successfully used to design many new biotechnologically valuable enzymes [Sen et al., 2007].

The role of complex methods of mutation vs. simple point mutation is currently an active area of discussion (e.g. [Long et al., 2003]). In particular, agents such as retroviruses (e.g. HIV) and retroposons are believed to work as highly effective and highly specific mutators (e.g. [Brosius, 1999]). The crossover mechanism evolved by retroviruses (Fig.2) shares many similarities with the DNA shuffling/sexual PCR techniques used in in vitro evolution.

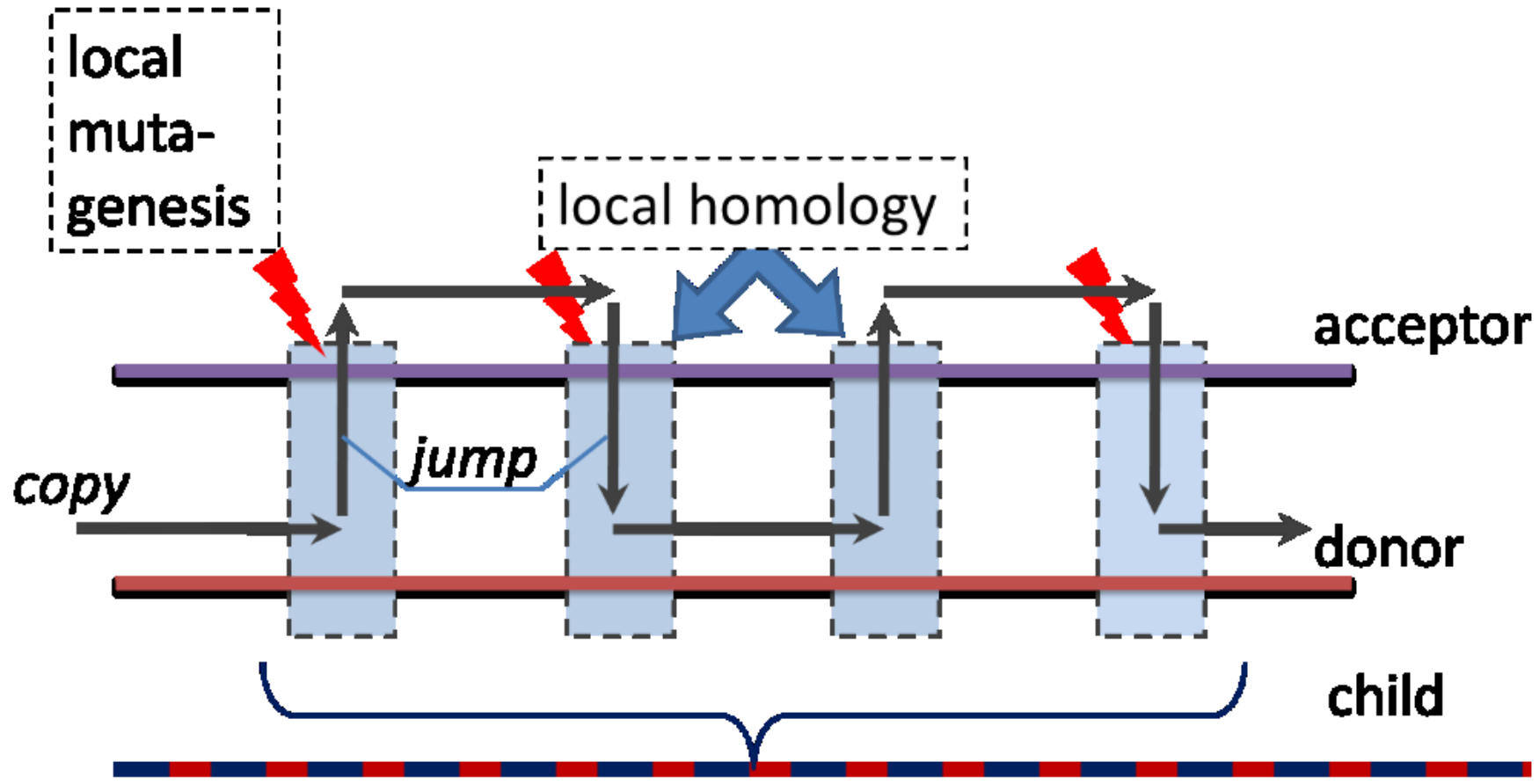

A

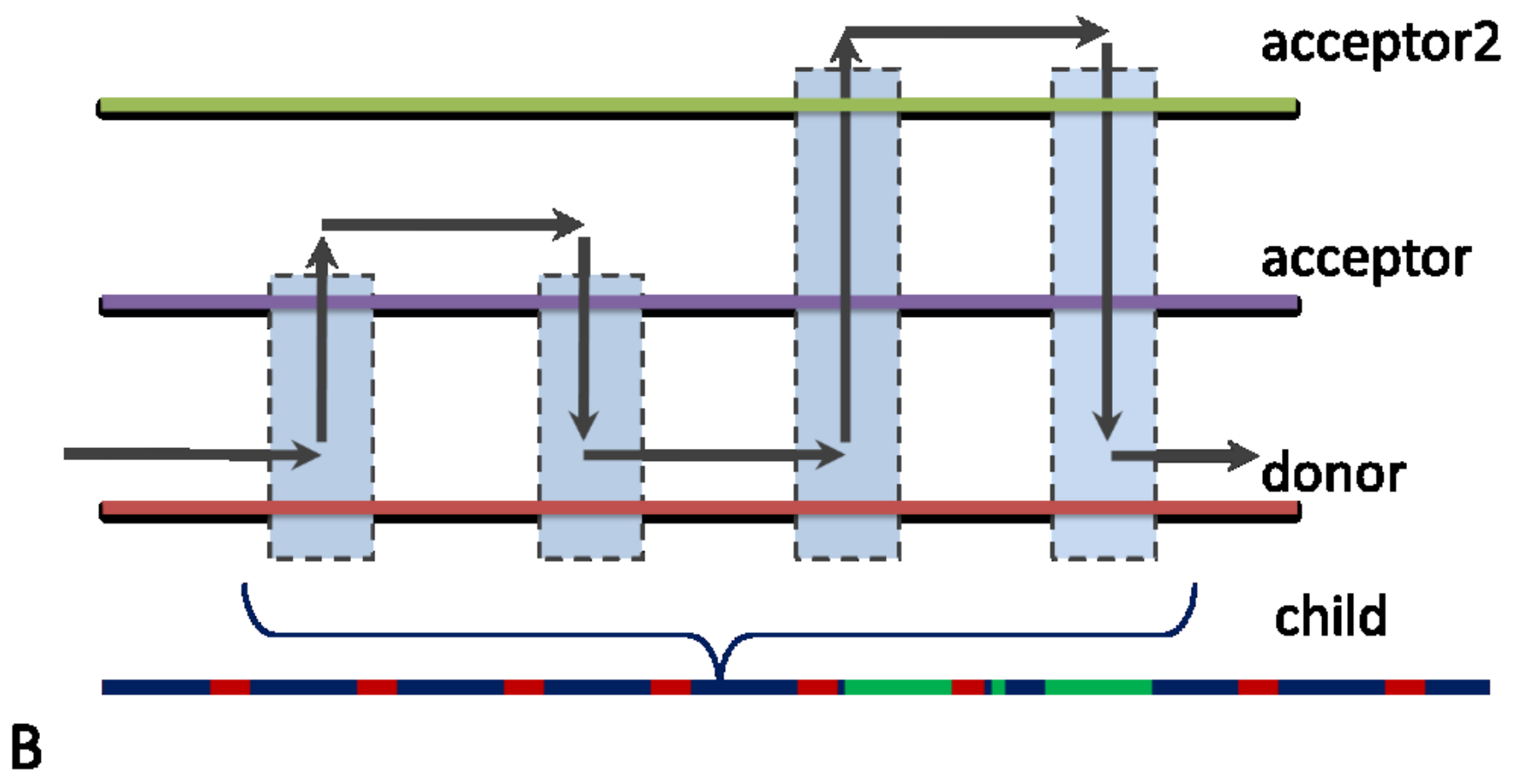

Figure 2. The overall idea of recombination between parental RNA strings used by retroviruses. (A) Regular case of homologous recombination. The child sequence is read alternately (horizontal arrows) from the parent strings, jumping between the parent templates (vertical arrows) at regions of homology (marked by gray rectangles). (B) Rare (still hypothetical) case of the three-string recombination (three “parents”: donor and acceptor and acceptor2).

Retroviral recombination usually takes two parent RNA strands to create a child DNA strand, though a three strand mechanism is also a possibility [Negroni, Buc, 2001; An, Telesnitsky, 2002]. The child sequence is read alternately from the parent strings, jumping between the parent templates at regions of homology (Fig.2). This is an effective mechanism for genetic diversity in the child, while retaining BBs. Development of GA crossover operators that use the retroviral scheme, or extend it to the multiple parent case seen in sexual PCR, can give quantitative understanding of the efficiencies of these techniques, and can provide insight into the biological evolution of retroviruses and retroposons.

# 3. Our Approach

We first consider a couple of tasks based on the problems of directed evolution of strong bacterial promoters like rrn1 (in comparison with the well-known RR & RS problems). This is a simplified rrn1 task and a more advanced one. Then, based on the results obtained and the conclusions reached, we study, by numerical experiments, the problem of finding an RNA device that is close to the problems of real life.

We will focus on our retroGA approach, using retroviral recombination methods (crossover) to preserve BBs during evolutionary searches.

## 3.1 GA benchmark functions as models of biomolecular evolution

John Holland’s schema theorem, also called the fundamental theorem of genetic algorithms [Holland, 1975], is an inequality that results from coarse-graining an equation for evolutionary dynamics. The Schema Theorem characterizes the behavior of short, low-order schemata with higher above-average fitness. It was initially widely taken to be the foundation for explanations of the power of GA.

A schema is a template that identifies a subset of strings with similarities at certain string positions. For example, consider binary strings of length 6. The schema 0*01*0 describes the set of all strings of length 6 with 0's at positions 1, 3 and 6 and a 1 at position 4. The * is a wildcard symbol, which means that positions 2 and 5 can have a value of either 1 or 0. The order of a schema is defined as the number of fixed positions in the template, while the defining length is the distance between the first and last specific positions. The order of 0*01*0 is 4 and its defining length is 5. The fitness of a schema is the average fitness of all strings matching the schema. The fitness of a string is a measure of the value of the encoded problem solution, as computed by a problem-specific evaluation function. Using the established methods and genetic operators of GA, the schema theorem states that short, low-order schemata with above-average fitness increase exponentially in successive generations.

Among the many benchmark tests in EC, the RR fitness functions were specifically invented to study the preservation and destruction of BBs by crossover operators. As such, they can serve as models for many cases of natural and test-tube evolution, in which searches proceed with BB preservation. Four RR functions, of increasing complexity, were invented and introduced by Forrest, Mitchell, and Holland to specifically test crossover operations in GA [Forrest, Mitchell, 1993a,b; Mitchell et al., 1992]. The related RS functions were devised and introduced by van Nimwegen and Crutchfield [2000; 2001]. These well-defined functions allow for analytical (mathematical) study of the evolutionary search behavior and parameter dependence.

### 3.1.1 Royal Road functions

Royal Road functions were devised to award fitness for the preservation of BBs, and thus to serve as models for natural evolution [van Nimwegen, Crutchfield, 2000]. R1, the simplest function, calculates bit string fitness by the number of order-8 schema, or words, in the string (from $s_1$ to $s_8$). Order does not matter:

```
s1 = 1111111**************************************************; c1 = 8
s2 = *******1111111*******************************************; c2 = 8
...............................................................................
s7 = *******************************************1111111*******; c7 = 8
s8 = **************************************************1111111; c8 = 8
sopt= 111111111111111111111111111111111111111111111111111111111; copt= 64
```

where the * is a random bit (0 or 1). The fitness value R1 (for string x) is the sum of the coefficients $c_s$ corresponding to each given schema of which x is an instance ($c_s$ is equal to order). The fitness of an intermediate step (such as the combination of $s_1$ and $s_8$) is a linear combination of the fitness of the lower level components (e.g. the combination of $s_1$ and $s_8$ has fitness 16). The genotype space consists of all bit-strings of length 64 and contains 9 neutral subbasins of fitness 0, 8, 16, 24, 32, 40, 48, 56 and 64. There is only one sequence with fitness 64, 255 strings with fitness 56, 65534 strings with fitness 48, etc. Because fitness proceeds by the build up of words, the fitness landscape for RRs has a subbasin-portal architecture, in which evolution tends to drift in neutral subbasins, with rare jumps to the next level via portals (creation or destruction of a word - schema).

In searching for fitness functions that are easy for GA and difficult for non-evolutionary methods, a whole family of increasingly complex RR functions was devised (R1, R2, R3, R4; [Mitchell, 1996]). These showed that standard GA is superior to non-evolutionary techniques on harder problems, but also brought to light that standard GA has substantial weaknesses in the standard crossover operators. Due to their formal simplicity, theoretical analysis can be carried out on the RR functions to understand the parameters which promote efficiencies.

R2 is very similar to R1, but allows for higher fitness at certain intermediate steps. R3 allows for random-bit spacers of set length between BBs (which are not calculated in the fitness score). The optimal string for R3 is:

$s_{opt}$=11111111********11111111********11111111*******11111111********11111111********11111111********11111111********11111111.

The most difficult RR function is R4, since one, two, or even four non-neighboring elementary (in our case 8-bit) BBs in the string gives the same exact score (Level 1). The score will only increase if words neighbor, e.g. a pair of juxtaposed 8-bit BBs creates a Level 2 16-bit BB. A Level 3 BB consists of 4 neighboring 8-bit BBs (32 bits in total). Level 4 BBs are 64-bit, composed of eight 8-bit elementary BBs. Level 5 BBs would consist of 16 8-bit BBs. Most current optimization techniques cannot effectively deal with the R4 fitness function.

From the viewpoint of molecular biology, R1 and especially R3 are reminiscent of a key element of the regulatory region of a gene: a cluster of binding sites (BS’s), made of short BBs separated by spacer sequences. While the analogy is good, there are several differences to bear in mind. First, usually the positions and order of BS’s in such clusters are less restricted (than in the comparable RR), but this depends on the particular gene in question (enhanceosomes vs. “billboards”; see [Jeziorska et al., 2009]). Second, any BS is not a unique sequence: it is usually a family of related sequences with varying strength (fitness), usually with a conserved core sequence [Stormo, 2000]. That is the BS can be seen as a particular case of a schema in the 4-letter alphabet. Finally, proximity of BSs to each other is important for the action of activators and repressors. This is analogous to R4 (e.g. with sub-clustering represented by R4, Level 3), but the biological spacing is somewhat less restricted than in R4 (see [Spirov and Holloway, 2012]).

### 3.1.2 Royal Staircase functions

The Royal Staircase class of fitness functions are a generalization of the RR functions in which the subbasin-portal architecture is expressed in a more explicit form [van Nimwegen, Crutchfield, 2000]. The RS functions depending on two parameters, the number N of blocks and the number K of bits per block, is defined as follows:

Genotypes are specified by binary strings , of length .

Starting from the first position, the number I(s) of consecutive 1s in a string is counted.

The fitness . The fitness is thus 1 plus the number of consecutive fully-set blocks (epochs) starting from the left.

The single global optimum is the string of all 1s.

The RS function similar to the 8-bit word, 64-bit string R1 and R2 of the previous section, but order matters (i.e. the string is built up from one end), and fitness for N=8, K=8 RS ranges from 0 to 80:

$s_1$ = 11111111********************************************************; $c_1$ = 10

$s_2$ = 1111111111111111************************************************; $c_2$ = 20

………………………………………………………………………………………………………………………………………………………………

$s_6$ = 111111111111111111111111111111111111111111111111****************; $c6$ = 60

$s_7$ = 11111111111111111111111111111111111111111111111111111111********; $c7$ = 70

$s_{opt}$= 1111111111111111111111111111111111111111111111111111111111111111; $copt$ = 80

Because of the general parallels between RR, RS and biological structure, we expect RR, RS analysis to shed some light on the evolution of gene regulatory regions, and to be useful as a theory for forced molecular evolution of bacterial and yeast gene promoters. (Where modified or completely artificial promoters can become new molecular tools for bio-sensing, etc. [Haseltine, Arnold, 2007; Lu et al., 2009; Keasling, 2012; Dehli et al., 2012; Berg et al., 2013]).

## 3.2 Our tests for directed evolution in silico

We consider here two simpler tests based on the sequences of strong bacterial promoters (simplified and extended rrn tests) and four-modular RNA device test.

### 3.2.1 Evolution of gene regulatory regions (benchmark tests)

Here we develop a prototypical case in the evolution of gene regulatory regions, highlighting their similarities to the RR and RS test functions.

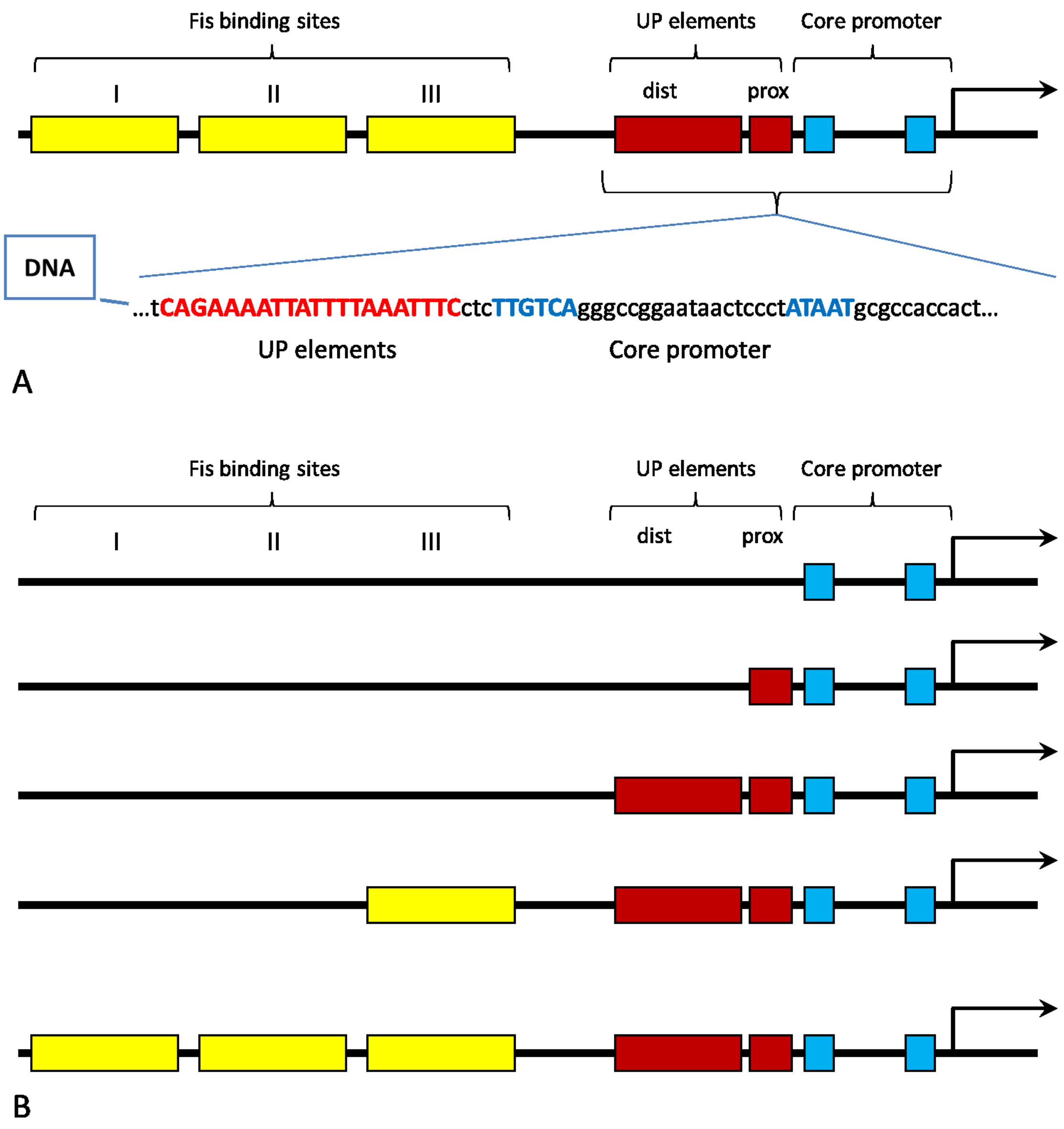

Figure 3. An example of molecular evolution similar with Royal Staircases. (A) Organization of the bacterial promoter rrnP1 (ribosomal RNA operon promoter) into a series of highly conserved blocks, with between-block spacers of conserved length. (B) Evolution of the rrnP1 promoter can be viewed as a Royal Staircase fitness function: starting with the core promoter, evolution could add the powerful UP element and then sequentially add FisBSs.

Bacterial gene promoters, being simpler than eukaryotic promoters, present good cases for investigating the details of the evolutionary searches producing their structure. For a target sequence (solution) for these sorts of problems, we have selected the sequence of the ribosomal RNA (rRNA) operon promoter rrnP1 in E. coli (Fig.3), since it is very well studied and well characterized [Schneider et al., 2003]. Core promoters in E. coli are approximately 60 base pairs (bp) long and are characterized by several conserved sites with spacers in between. It is believed that while the sequences of these spacers are not significant, their lengths are of extreme importance [Schneider et al., 2003]. There are at least four well-conserved features in a bacterial promoter: the starting point (usually 'CAT'); the -10 sequence ('TATAAT' consensus); the -35 sequence ('TTGACA' consensus); and the distance between the -10 and -35 sequences (Fig.4A). The rrnP1 promoter sequence contains an AT-rich sequence called the upstream (UP) element [Ross et al., 1998] upstream of the -35 element. UP elements increase transcription 20-to 50-fold [Hirvonen et al., 2001]. Its consensus is AAA a/t a/t T a/t TTTT**AAAA, where * indicates a random base, a/t means A or T. The UP element consists of two distinct subsites (that is, a proximal subsite, and a distal one; Fig.4B), each of which, by itself, can bind the RNA polymerase holoenzyme a subunit carboxy-terminal domain and stimulate transcription [Estrem et al., 1999]. In addition, three to five BS's for the Fis protein binding sites (FisBS; Fig.4C) increase transcription three- to eight-fold [Ross et al., 1990]. It is crucial, that mutations at rrn~~E~~ P1 that prevent binding of FIS to site I eliminate activation by distal FIS sites [Estrem et al., 1999]. The weight matrix for the binding sites of this transcription factor has been defined [Hengen et al., 1997]. The desired sequence for the rrnP1 promoter therefore includes:

**[FisBS]**<~5 bp>**[FisBS]**<~5 bp>**[FisBS]**<~15 bp>**AAA a/t a/t T a/t TTTT**AAAA**<~4 bp>**TTGACA**<16-19 bp>**TATAAT**<5-9 bp>**CAT.**

Hence, evolution of the rrnP1 promoter can be imagined as follows (as presented in Fig.3). The core bacterial promoter is capable to keep gene expression, but at low level. The core promoter strengthened by proximal (then distal) UP element is able to maintain gene expression at substantially higher level. Addition distally the first FisBS enlarges the promoter strength even higher. Finally, adding several more FisBS strengthen the promoter even more.

It is promising for our consideration that practically all the promoter cases are observable in prokaryotic genomes and can be reproduced in lab (e.g. [Ross et al., 1998; Estrem et al., 1999; Hirvonen et al., 2001; Schneider et al., 2003]). Hence we can hypothesize that this is the way how one of the strongest prokaryotic promoters, rrns, could evolve.

Therefore, evolution of the rrnP1 promoter can be viewed as a RS fitness function. Starting with the core promoter ($s_1$), evolution could add the powerful UP element ($s_2$) and then sequentially add FisBSs ($s_3$, $s_4$):

```
s1=*********************************…************************************************TTGACA***...***TATAAT***...***CAT, c1 = Δ
s2=***************…*****************************AAA a/t a/t T a/t TTTT**AAAA***...***TTGACA***...***TATAAT***...***CAT, c2 = ~35 Δ
s3=************…*****************[FisBS]*** …***AAA a/t a/t T a/t TTTT**AAAA***...***TTGACA***...***TATAAT***...***CAT, c3 = ~100 Δ
......................................................................................................................................
sopt=***[FisBS]***…***[FisBS]***…***[FisBS]*** …***AAA a/t a/t T a/t TTTT**AAAA***...***TTGACA***...***TATAAT***...***CAT, copt = ~150Δ
```

where Δ is an arbitrary small fitness value.

Like RS, rrnP1 probably evolved by sequential finding and adding of BBs, with each addition raising the fitness of the promoter sequence (transcriptional efficiency). As with RS, the length and positions of the BBs are conserved during evolution, though not as strictly as the RS function.

*3.2.1.2 The rrnP1 benchmark-tests*

It is well known that functional clusters of BBs in biological macromolecules display the redundant character of the blocks. Functionally very similar blocks can have different sequences, sharing only a common core sequence. Besides, they are not binary, but quaternary (four-letter DNA and RNA sequences). In fact, BBs here are schemata in 4-letter strings, with more complicated organization comparing with binary strings: some positions can be fixed (A only, etc.), some positions are redundant ("*" positions), but the rest can be semi-redundant (say, a/t means either A or T, but not G or C, a/t/c means either A or T or C, but not G, etc.). The cases of such schemata for our test-problem are illustrated in Figure 4. Also, compared to RR and RS, biological clusters of BBs include longer spacers (which is often of variable length), and they are usually longer than 64 or 128 elements.

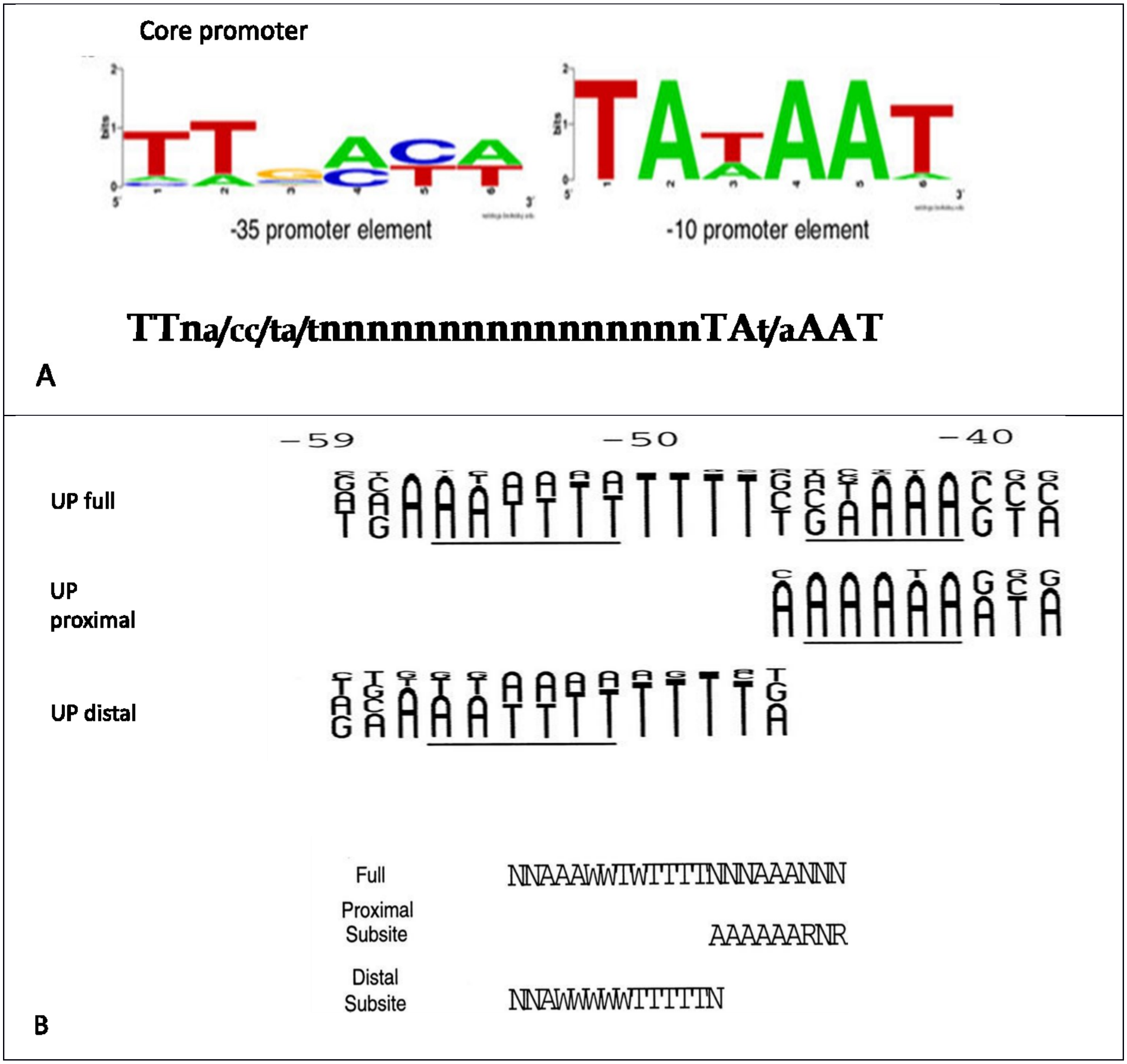

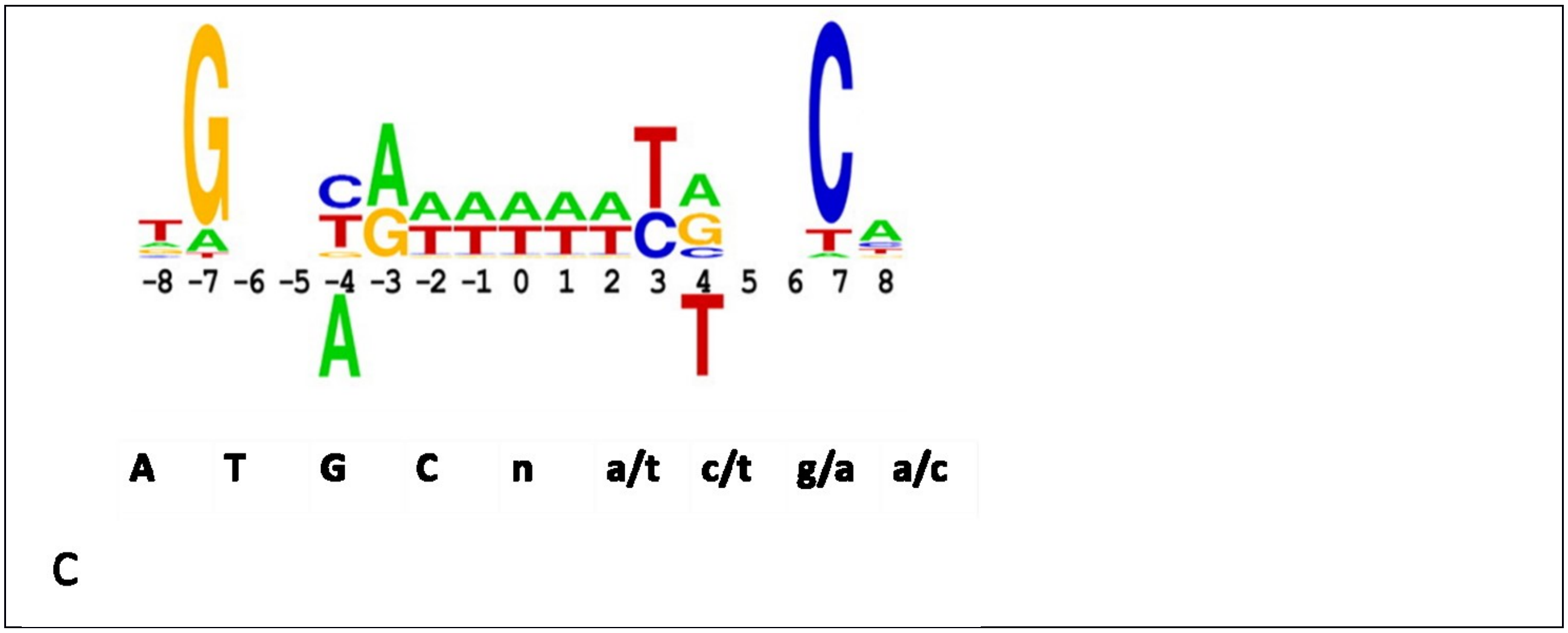

Figure 4. The consensus sequences and consensus logo for the ribosomal RNA (rRNA) operon promoter rrnP1 in E. coli. These are the core promoter sequences (A), the upstream (UP) element, proximal and distal sequences (B), and the Fis protein binding sites (C).

In our consideration, all schema cases of a given type (e.g. all FisBS, all UP proximal elements, etc.) are characterized, for simplicity's sake, by the same fitness value (e.g. 1 for the core promoter consensus, 4 for the core promoter plus the UP proximal consensus, 36 for core promoter plus complete UP consensus, etc).

The simplified rrnP1: To begin studying the rrnP1 problem (and do so within the RR, RS framework), we have simplified some of these complications: we ignored the redundant character of its 8 BBs and the variability of spacer lengths; we assumed that all the elements are fixed and/or unique in sequence; and we considered five elements only. The first of these represents the whole core promoter and was modelled by only 6 letters. The second element was the proximal half of the UP element, assumed to have a length of 5 letters. The spacer between the 1st and the 2nd elements was 24 letters. The 3rd to 5th elements are given the same length {=5}, with spacers of 15-letter length. We will call the rrnP1 case as the *simplified* one. The results on computing this simplified target were presented in [Spirov and Holloway, 2012] and will be compared with this paper results.

The extended rrnP1: In this paper we develop and explore more realistic and computationally harder case of the rrnP1 "benchmark"-test. Here we put into consideration the redundant character of its 8 BBs (Fig.4), but keep ignoring the variability of spacer lengths (that is the positions and the sizes of all the elements are fixed). We consider four elements only: the core promoter, UP proximal, UP distal, Fis1 and Fis2. We will call the test case as the *extended* one.

### 3.2.2 The RNA-device tests (real-life tests)

In order to get closer to the problems of real life, we also formulated tests that resemble the tasks of the directed evolution of RNA devices. The main simplification used by us is dictated by the very large computational cost of evolutionary tests for RNA sequences with dimensions comparable to real RNA devices. The most advanced and studied RNA devices of only 3 (- 4x) modules have a length of a few hundred nucleotides (e.g. [Koizumi et al., 1999; Koizumi, Breaker, 2000]).

Here we consider a test consisting of four RNA domains belonging to 4 classes of small RNAs, such that they correspond in length to the shorter of the known sequences. Specifically, we will select 5-nucleotide long domains (GUGGC) for the minimum five-nucleotide ribozyme, minR; 11-nucleotide long domains (M4U:

AAAA. . . AGUAGUC), for M4U; 18-nucleotide long domains (hsa-miR-516a-3p: UGCUUCCUUUCAGAGGGU), for Hmir; and 19-nucleotide long domains (Inteleukin-6 eceto: GGGGAGGCUGUGGUGAGGG), for IntL, separated by structural sequences from 15 to 5 N. The functional module, composed of parts separated at fixed distances, is treated as one domain with adjacent spacers.

In order to bring our test closer to the problems of real life and, at the same time, reduce its computational cost, we consider optimal target sequences to be consensus and accept a rather high degree of sequence degeneracy for each of the domains. We have chosen a measure of the degeneracy of domain sequences so that the price of each of the 4 domains is approximately the same.

Evolution of the RNA device can be viewed as a RS fitness function: starting with the sequence with the ribozyme minR, evolution could add the M4U domain, then add the hsa-miR-516a-3p domain, and finally add the Inteleukin-6 eceto domain.

Moreover, for greater realism, we are not looking for each domain in its unique, unambiguous position on the RNA sequence. In contrast, each of the domains can be found within a window of 10 nucleotides. That is, freedom of localization of the desired motive is set over the course of M + 10, where M is the length of the motive itself.

## 3.3 Approaches Preserving Already Found Blocks

As mentioned in the Introduction, standard GA techniques, specifically through the use of crossover to generate diversity in the chromosome, can destroy BB's which are important for fitness, slowing evolutionary searches. Among the variety of approaches that preserve the blocks already found, we selected only a few in order to demonstrate the prospects of their implementation in real experimental techniques of in vitro evolution.

**The HRO and the retroGA technique**: We have taken inspiration from the mechanisms of retroviral recombination to create crossover operators which preserve BB's. Our innovations are only in the crossover operators, all other actions of the algorithm are as in classical GA. We note here that as a long-standing predecessor of our approach, we should mention the homologous recombination approach [Park et al., 1993]. In order to keep the information and to make the recombination possible, HRO considers only restricted region which shows homology over a specified threshold value, when it selects crossover points. This approach as the simplest implementation of our techniques will be tested in our article.

As discussed above, homology-based PCR techniques (DNA shuffling, sexual PCR) used in test tube evolution may be naturally interpreted as a generalization of retroviral recombination processes (Fig.2), using n instead of 2 parent strings. Our retroGA operator generates a child string from a given "parent set", combining the function of reproduction and crossover. Crossover points are determined by regions of local homology in the parent strings. The parent strings are selected from the population, as in standard GA, by one of several predetermined strategies, such as truncation, roulette-wheel, etc. One string is selected as a donor, the others as acceptors (Fig.5).

In our reproduction and crossover procedure, a first pair of parent candidates is selected. These are the donor and acceptor-1 (Fig.5). Their sequences are then compared going from left to right for a short distance $L_{acc}$ (where $L_{acc} < L$, L is the length of the whole sequence). If the required zone of local homology is not found, another candidate for acceptor-1 is selected. The number of attempts to find a suitable acceptor is at most $N_{acc}$. If, and only if, a zone of complete homology of a size no less than $L_{hom}$ symbols ($L_{hom} < L$) is found during an attempt to scan two sequences, do these two sequences become the donor and acceptor-1 pair. Replica generation is then initiated, and takes place in the first n symbols of the donor, from the first element to the last element of the region homologous between the two parents. Replication then jumps to

acceptor-1, and acceptor-2 candidates are selected. A search for local homology takes place between acceptor-1 and the putative acceptor-2. If no such region is found, the next candidate is searched. This process is iteratively repeated until the replica (child) is completed, or until the $N_{acc}$ limit is exceeded.

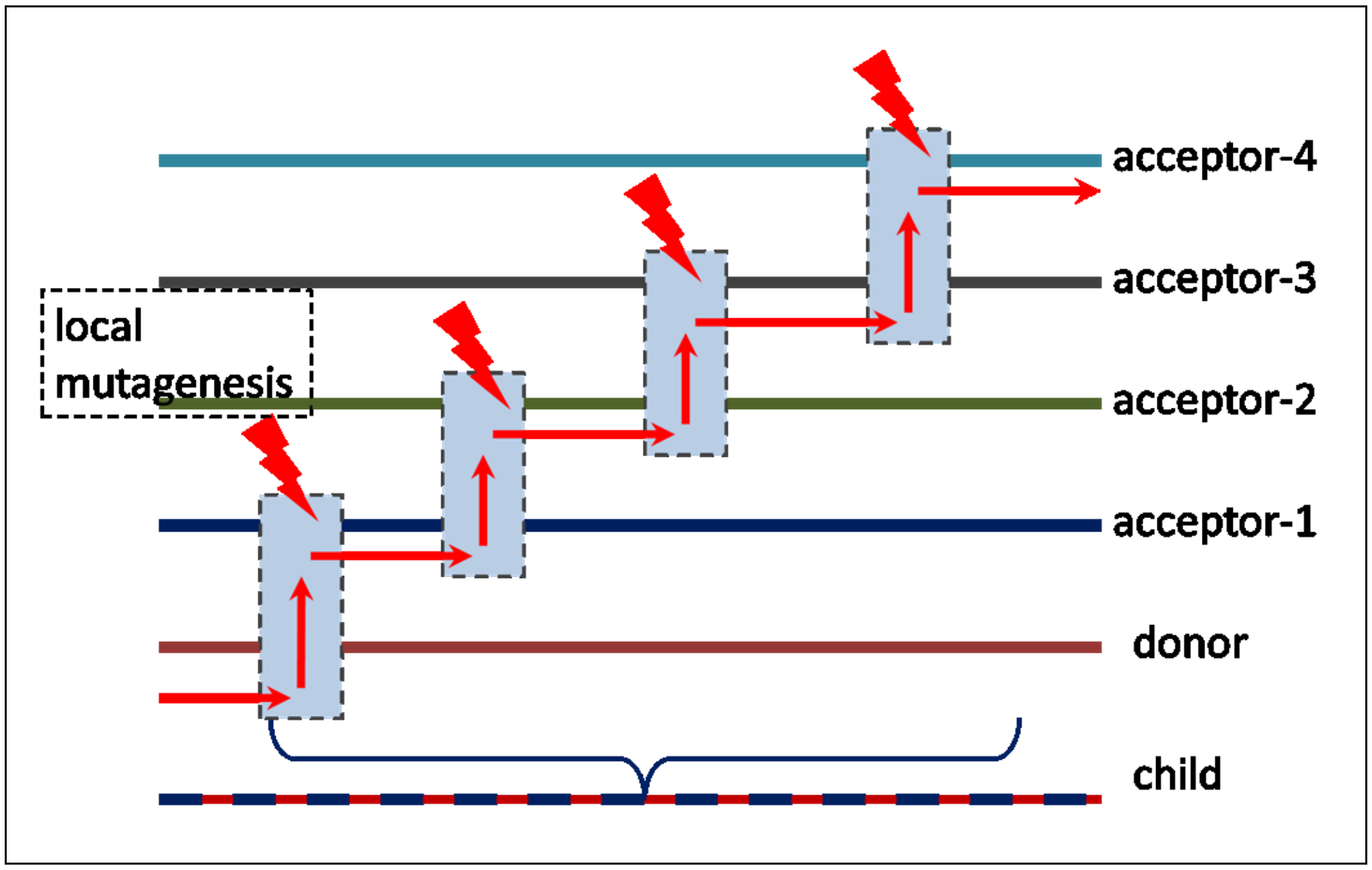

Figure 5. Principle of the retroGA operator, an analogy to retroviral recombination and in vitro DNA shuffling techniques. The process of creating the child sequence by the operator starts with the donor-parent. Replication of the child from the donor-parent occurs if there is at least one region of homology (identity) between the donor and acceptor -1 (marked by gray rectangles). The process then jumps onto the acceptor-1 string. An acceptor-2 is found with a region of homology to acceptor-1, and the process repeats, copying from acceptor -1 and jumping to acceptor-2 (which becomes the third parent of the child sequence). The process of jumping from acceptor to acceptor continues until the creation of the child sequence is complete.

*retroGA with point mutations*: In real retroviral recombination it appears that both crossover and point mutations are present. Template switching between parent RNA strings tends to introduce mutations in the child sequence. For our retroGA, we include this effect by introducing one point mutation in one of a few starting sites in the portion of the child string being copied from the new acceptor. This addition provided speed-up for retroGA on RS, and rrnP1-gene searches, but not on RR searches [Spirov and Holloway, 2012].

We emphasize here that the choice of specific algorithms, as well as specific values of their parameters, was made by us from those key considerations so that our models are as close as possible to the experimental evolution in vitro. This is, first of all, a rigidly set percentage of mutations when copying offspring (from parents); an algorithm-rule for selecting parents (fitness threshold); Typically, there is no crossover in the experiment.

**Random-mutation hill-climbing (RMHC)**: The algorithm, proposed back in the early 90s, is similar to a zero−temperature Metropolis method [Forrest, Mitchell, 1993a;b; Mitchell, 1996]. The procedure is as follows:

1. Choose a string at random. Call this string best-evaluated.

2. Choose a locus at random to mutate. If the mutation leads to an equal or higher fitness, then set best-evaluated to the resulting string.

3. Go to step 2.

4. When a set number of function evaluations has been performed, return the current value of best-evaluated.

RMHC found the optimum on R1 nearly a factor of 10 faster than the GA [Mitchell, 1996].

**Other approaches preserving BB**: Some crossover (or crossover-related) algorithms capable to protect BB, were introduced in EC recent decades (e.g. [Yu et al., 2009; Kameya, Prayoonsri, 2011; Sanchez et al., 2017]). Particularly, linkage learning methods have been developed intensively, and these linkage learning methods basically assume a fixed-position (binary) encoding of chromosomes and BBs are identified/protected in a locus-wise fashion.

*Table 1 Algorithms preserving BBs*

| Algorithm name | Reference | Comments |
|---|---|---|
| GAP (GA with patterns) | | |
| HRO (homologous recombination operator) | | |
| Puzzle Algorithm | | |
| Co-Puzzle Algorithm | | |
| Rough Set theory, RSO algorithm | | |
| Statistics-based Adaptive Non-uniform Crossover (SANUX) | | |
| Looseness-controlled Crossover (LCC) operator | | |
| Relevant Alleles Preserving GA (RAPGA) | | |
| Subset Size Oriented Common Features (SSOCF) operator | | |
| Self-Adaptive Semantic Crossover, (SASC) | | |

Here we will use only the basic idea of protecting already found blocks. Since the procedures discussed in our article are focused on the prospects of applications to real-life problems, the problems are formulated so that we usually know the localization of the already found block. Therefore, we can exclude it from the scope of mutations (and crossover).

# 4. Results

In this section we present results on the comparisons of the introduced test-problems (the RR & RS problems vs rrnP1 and RNA-device tests), as well as on efficiency of retroGA in comparison with standard GA. Because all of these problems share a subbasin-portal type architecture, such computations allow us to begin to characterize the degree to which RS test functions can predict behavior in the gene design searches. This is especially relevant if we can begin to use the analytical (mathematical) tools that have been developed for the RR & RS test functions to understand the gene search dynamics.

## 4.1 Crossover operators for RR & RS benchmark problems

We began our numerical experiments with our test problems by analyzing how average time to achieve a given fitness n empirically depends on n.

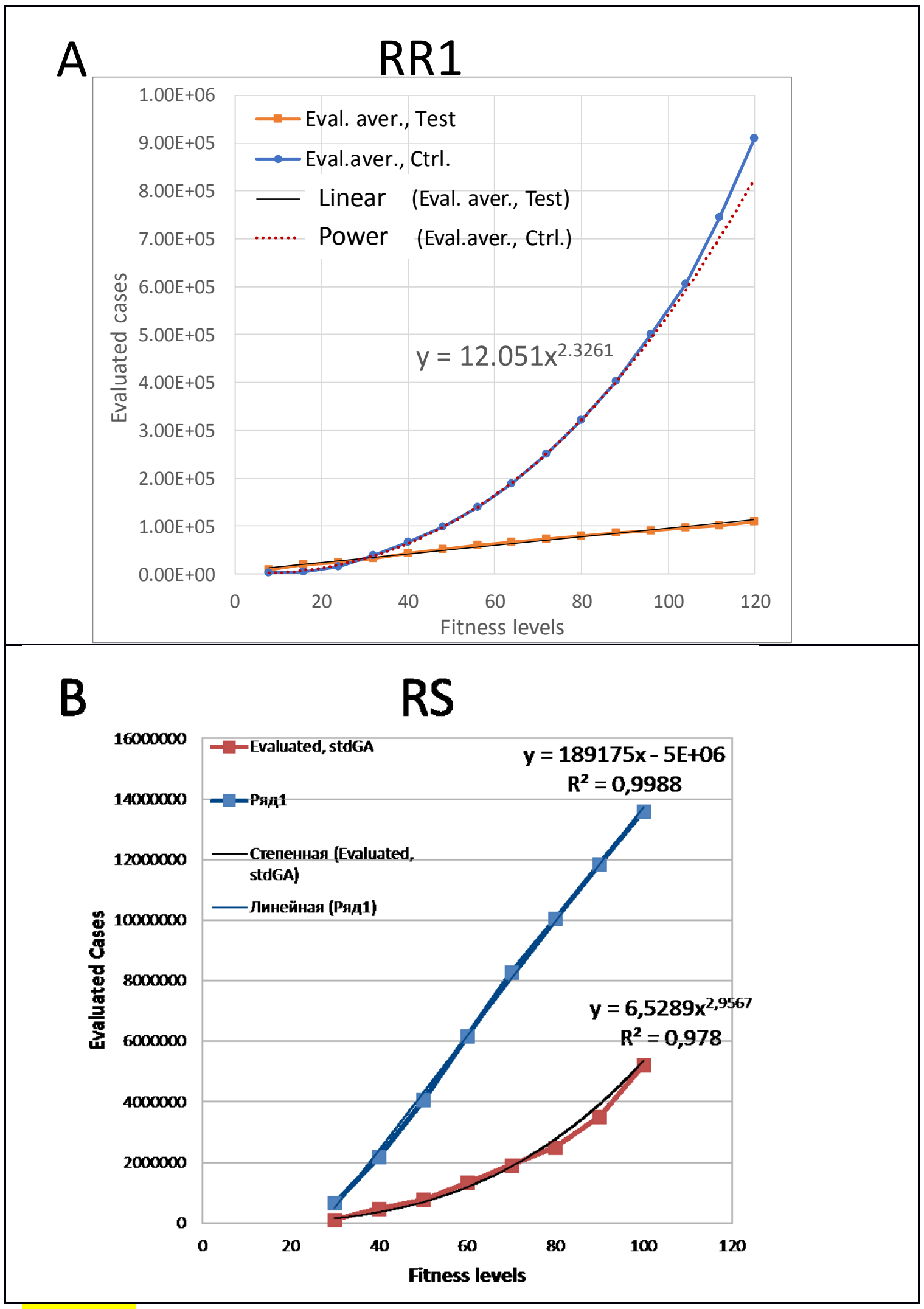

Figure 6. This is how average time to achieve a given fitness n empirically depends on n for RR and RS tests. (A) Average total number of fitness function evaluations to achieve nth fitness level for the Royal Road 1 with N=16 blocks of length K=8. Evolutionary search with point mutations only vs. search by retroGA algorithms. Each data point was obtained as an average over 200 GA runs. Population size M=3,000; mutation rate q=0.01, truncation selection strategy. (B) Average total number of fitness function evaluations to achieve nth fitness level for the Royal Staircase with N=10 blocks of length K=10. Evolutionary search by retroGA algorithms (B) vs search with point mutations only. (C). Each data point was obtained as an average over 200 GA runs.

We have already reported on the several-fold speed-up of retroGA vs. standard GA for RR and RS problems [Spirov & Holloway, 2010; Spirov & Holloway, 2012].

Here we further characterize retroGA performance on RR test (R1). These functions have been well-studied in the literature, and as discussed above, have some of the fundamental motifs necessary for modeling gene organization. We have already reported on the several-fold speed-up of retroGA vs. standard GA for RR problems [Spirov & Holloway, 2010]. Here we will focus on how average time to achieve a given fitness n empirically depends on n for higher N=16, K=8, (Fig.6). For the R fitness function the solution by the retroGA crossover/point mutation was achieved about 5.5 times faster than by standard GA (point mutations only).

In developing a theory for the R1 problem, van Nimwegen and colleagues [1999], predicted that epoch duration depends exponentially on epoch number (fitness level) n. Computationally, we do see a roughly nearly exponential dependence for standard GA (no crossover; Fig.6A), though the empirical dependence is well approximated by a power function also; See Fig.6A). Our retroGA shows a linear relationship between number of evaluations and n for R1 (Fig.6A).

The results with the RS of N = 10, K = 10 is shown in Fig.6B. Testing our retroGA operators clearly shows that the combined crossover/point mutation mechanism is the most effective procedure on RS tests, speeding up searches substantially [Spirov and Holloway, 2012]. For the RS fitness function N=**10**, K=10, the RS optimum by the retroGA crossover/point mutation was achieved about 5.5 times faster than by standard GA (point mutations only) [Spirov and Holloway, 2012].

In [Spirov and Holloway, 2012], we have corroborated the RS results of van Nimwegen & Crutchfield [2000; 2001] with point mutation GA. Here we observe that the averaged time (in the average number of candidate string evaluations) to achieve the n+1 fitness level rises nearly exponentially (but the empirical dependence is well approximated by a power function also; See Fig.6B).

The effectiveness of retroGA certainly depends on the parameters and a thorough study of these dependencies on the example of the RR problem has already been published by us [Spirov and Holloway, 2012].

## 4.2 Crossover operators for rrn problems

We previously found that the simplified version of the rrnP1 test behaved very closely to the RS tests with N=4, K=10 [Spirov and Holloway, 2012]. The dependence of the search efficacy on the n level is approximately exponential (Fig.7A) for retroGA on rrnP1, as on RS. retroGA on the simplified rrnP1 (with five blocks) was over five times faster than GA with one-point crossover [Spirov and Holloway, 2012].

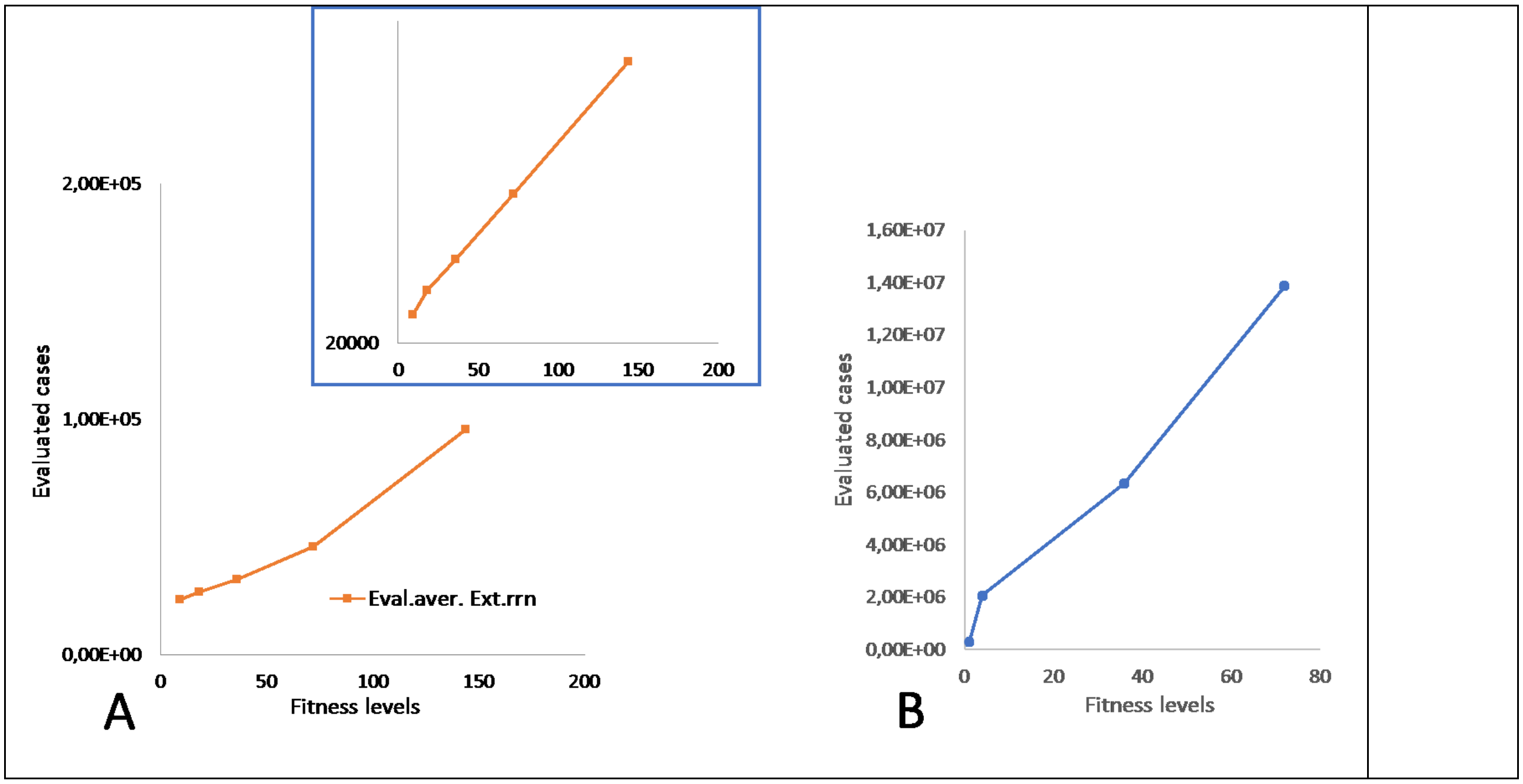

Figure 7. Epoch durations for the rrn problem (simplified, A, and extended, B): time to achieve fitness level n by retroGA.

Our current tests on the extended rrnP1 benchmark task of the project reveal even higher efficacy for retroGA (with mutations) vs standard GA. retroGA on the extended rrnP1 test achieves the last, 4$^{th}$ level in 13,591,932± 1,751,020 (average over 100 runs; Fig.7). To our surprise, standard GA practically unable to achieve the 4$^{th}$ level: the success rate was a few per-cents only.

As we can see in the Fig.7 plots, the extended rrnP1 test is substantially more time-consuming than the simplified one. There is a large price to pay for increasing the dimension of the problem.

Success on the rrnP1 problem previously and here, and the parallels to the well-characterized RS function, suggests that retroGA is an effective technique for prokaryotic gene search problems, and could contribute to real problems of forced (directed) evolution.

Our prospective goal is to use the techniques developed in this paper to aid the directed evolution of bacterial and yeast gene promoters in the laboratory. Several approaches to improve and/or analyze such promoters via directed evolution have been undertaken by experimentalists [Schmidt-Dannert, 2001; Haseltine, Arnold, 2007; Collins et al., 2006; Gilman, Love, 2016; Zhou et al., 2017]. While there is still some gap between the gene models in this paper and real macromolecular evolution, we hope to have outlined the directions that can be taken for the computational work to provide a stronger theoretical basis for directing and analyzing experiments.

## 4.3 Crossover to evolve RNA-devices

We performed computer experiments simulating the directed evolution of biological macromolecules (on the example of RNA devices). We demonstrate how the implementation of GA approaches to preservation of BB significantly increases the efficiency of tuning and modification of RNA devices. Specifically, we test HRO (Park with co-authors [1993]), and our retroGA approach, in comparison with standard mutation operator, and standard crossover & mutation, and compared to the RMHC algorithm.

We performed testing on three computational problems: evolutionary search for a single module (I); tuning of modular device (II); search for a new modular device from scratch (III).

In real experiments on the directed evolution of RNA devices, recombination-type procedures are not used at all, but only random mutagenesis, usually via error-prone PCR. In this section, we want to evaluate how procedures such as recombination can increase the efficiency of evolutionary search.

### 4.2.1 RNA-devices: modification and tuning

Tuning and modification of the device are the main approaches in the field of rational design of RNA devices. Here we consider the results of test tasks for modifying and tuning that RNA device.

Tuning problem: Here available at hand modules are stitched together into a new molecule, usually poorly functioning or not functioning at all. Then they use the approaches of experimental evolution to obtain a device capable of working.

Here we take a non-functional block of three modules, similar to the previous one, assembled by rational design from the available RNA elements. In this test, we investigate how to get a working device from this «workpiece» faster, i.e. we investigate the tuning problem. Specifically, we repeat in silico real experiments of the Smolke' laboratory [Liang et al., 2011; Kim, Smolke, 2017].

The initial population of the three-domain version of the RNA device is different from the target version in the 5$^{th}$ nucleotides for the 1$^{st}$ and the 2$^{nd}$ domain 5th and in 6$^{th}$ nucleotides for the 3$^{rd}$ one. In the test, the finding of the complete sequence of the sensor-aptamer is evaluated as the achieving the first level of fitness (C1). Finding the complete sequence of actuator (hammerhead ribozyme) gives C2 (C2>C1). Finally, the finding of the complete transmitter sequence gives a fully functional device with the fitness level C3 (C3>C2).

Our results demonstrate that retroGA provides acceptable performance for this, the most difficult task in our test problems. This is despite the fact that standard GAs generally cannot find the desired 3 domain consensus.

For our tests, we performed a series of runs (100 for tests 1-2 and 50 for test 3) with standard single- and two-point crossover operators, and the retroGA operator. The sizes of populations were chosen such as to achieve a high efficiency of evolutionary search. For test 1, the population size of 80 thousand individuals was sufficient. As a control, we used runs with standard GA operators (mutation and crossovers).

*Table 2 RNA-devices: tuning in silico*

| Control/Test | Efficacy, % | Evaluated ± Std.Dev. |
|---|---|---|
| Mutations, pop=480 тыс.; Qu-0,3. | - | - |
| Stdt crossover, pop=480 тыс.  Crs-0,1; Qu-0,3 | A few per-cent | |
| RetroGA, pop=160 тыс. HIVlim=8, Crs-0,1; Qu-0,3; _RpStp=GenLen div 4; RComb=GenLen div 3 | 60 | 14.978.594,43±3.273.309,15 |

*Table 3 RNA-devices: tuning in silico (2)*

| Test/Control | Initial mutation level | Efficacy, % | Evaluated±Std.Dev. |
|---|---|---|---|
| RetroGA, pop=60 тыс. HIVlim=8, Crs-0,1; Qu-0,3; _RpStp=GenLen div 4; RComb=GenLen div 3 | --<br>Reprod:Random(4) | 72 | **8066204.3±3918720.9** |
| -“- | IniPop:Random(12)<br>Reprod:Random(12) | 50 | 8619689.4±2578814.9 |
| -“- | IniPop:Random(24) | 60 | **6722376.5±2460324.7** |
| -“- | IniPop:Random(48) | 80 | **7234892.4±2956363.4** |
| -“- | IniPop:Random(48)<br>Reprod:Random(12) | 60 | 8450343.7±3771613.2 |
| -“- | IniPop:?? | 40 | 8381370.3±1366448.4 |
| Standard GA, pop=60 тыс. HIVlim=8, Crs-0,1; Qu-0,3; _RpStp=GenLen div 4; RComb=GenLen div 3 | IniPop:Random(48) | 10 | 5352941 |
| Standard GA, pop=320 тыс. HIVlim=8, Crs-0,1; Qu-0,3; _RpStp=GenLen div 4; RComb=GenLen div 3 | IniPop:Random(48) | -- | -- |

As a result, we come to the conclusion that the main problem in the tuning task is the level of diversity of the initial population.

Modification: We will take Koizumi with co-authors [Koizumi et al., 1999] experiments as a basis. Here is given the original functional three-module device (sensor, transmitter and effector-actuator), which is modified by experimental selection so that the sensor (aptamer) is eventually replaced by another one. That is, this is the problem of evolutionary modification of the available device.

Specifically, in the work of Koizumi with co-authors, the aptamer for the cAMP ligand molecule was selected from a 25-nucleotide randomized sequence in place of the sensor module. For more specificity of the test, we used the data on the functional organization of families of such aptamers, selected and analyzed in the subsequent publication by these authors [Koizumi, Breaker, 2000]. Namely, the stretch of 16 conservative nucleotides is necessary for the specific binding of the ligand, while the remaining nucleotides are less conservative and are necessary for the desired specificity and tuning of the aptamer.

*Table 4 Tests with modification*

| Control/Test | Evaluated ± Std.Dev. | Efficacy, % |
|---|---|---|
| CtrAMP: Popul=80,000; Limit=24000000; Quote=0.30; CrossGA=0.0100 | 3.626.147,85±4.284.054,57 | 80 |
| RetroGA: pop=70,000; Crs-0,1; Qu-0,3; | 4.245.380,64±3.165.995,15 | 88 |
| RetroGA: pop=160,000; Crs-0,1; Qu-0,3; HIVlim=8; _RpStp=GenLen div 4; RComb=GenLen div 3 | 7.488.490,85±5.487.021,69 | 84 |

For our tests, we performed a series of runs (100 for the tests) with standard single- and two-point crossover operators, and the retroGA operator. The sizes of populations were chosen such as to achieve a high efficiency of evolutionary search. For test 2 the population size was 60 thousand individuals (160 thousand for controls). As a control, we used runs with standard GA operators (mutation and crossovers).

As a result (Table 3), we conclude that the selection from scratch for a single domain by homologous (retroGA) recombination is not better than standard GAs.

### 4.2.2 RNA-devices: evolutionary design de novo

Finally, as a natural extension of the approach, we consider the problem of finding the desired device (as in the previous test), but we set it as the task to find a three-modular RNA device from scratch (from the initial population of random sequences of fixed length), as illustrated by Fig.1. We begin with the initial population of random sequences by selecting for the domain of the aptamer. Next, the population of the selected sequences we are subjected to selection for the second module – hammerhead ribozyme domain. Finally, we select the obtained population of two-domain sequences for a fully functional device.

Our comparison of tests with mutations and the standard crossover versus tests with point mutations only showed significantly higher efficiency of evolutionary search with crossover. The results are summarized in table 5. Due to the obvious resource intensity and difficulty of this test problem, we present three results for different population sizes (160, 320, and 480 thousand).

*Table 5 RNA-devices: evolutionary design de novo*

| Test | Efficacy, % | Evaluated± Std.Dev. |
|---|---|---|
| Mutations, pop=160 тыс. | 12 | 15.364.316,75±4746742,33 |
| Mutations, pop=320 тыс. Crs-0,1; Qu-0,3 | 64 | 9.401.010,19±3537608,52 |
| Stdt crossover, pop=160 тыс. | *64* | 5.983.121,31±3101212,27 |
| Stdt crossover, pop=320 тыс. Crs-0,1; Qu-0,3 | 63 | 10.376.104,75±4851423,62 |
| Stdt crossover, pop=320 тыс. Crs-0,2; Qu-0,4 | 65 | 11.768.089,44±2807074,38 |
| Stdt crossover, pop=480 тыс. Crs-0,1; Qu-0,3 | 72 | 13.008.087,04±4234476,38 |
| RetroGA, pop=60 тыс. Crs-0,1; Qu-0,3; HIVlim=8; _RpStp=GenLen div 4; RComb=GenLen div 3 | *96* | 2.695.806±349.188,08 |
| RetroGA, pop=75 тыс. Crs-0,1; Qu-0,3; HIVlim=8; _RpStp=GenLen div 4; RComb=GenLen div 3 | *98* | 3.257.485,19±474.771,1 |
| RetroGA, pop=80 тыс. Crs-0,1; Qu-0,3; HIVlim=8; _RpStp=GenLen div 4; RComb=GenLen div 3 | 100 | 3.545.264,38±612.328,18 |

As a result, we found that the recombination algorithms that preserve BB (retroGA) are significantly more efficient than the mutation only and more efficient than the standard mutation + crossover. The efficiency increases with the number of required modules. For the modification problem (test 1), standard GA algorithms were comparable to the retroGA. For the task of tuning a three-domain device (test 2), the retroGA procedure is substantially more efficient than the standard mutation operator. Our algorithm requires the evaluation of 7.234.892,4±2.956.363,4 individuals in average to find the desired solution (with an efficiency of 76%). While the efficiency of the standard algorithms was extremely low - a few percent.

For the tasks of searching for multi-modular RNA devices from scratch (test 3) the effectiveness of non-trivial heuristic recombination algorithms is impressive also. retroGA finds the desired sequence of the three-modular device with high efficiency (7.545.264,4±2.612.328,2 evaluated, 80% efficacy), while the standard GA shows low efficacy even for populations of 480 thousand.

### 4.2.3 RNA-devices: Time to achieve a given fitness n

It is curious that in this task of sequentially searching 4 domains (IntL = 19, Hmir = 18, M4U = 11 and minR = 5) with the retroGA algorithm, the speed of finding each next block is almost linear (Fig.8).

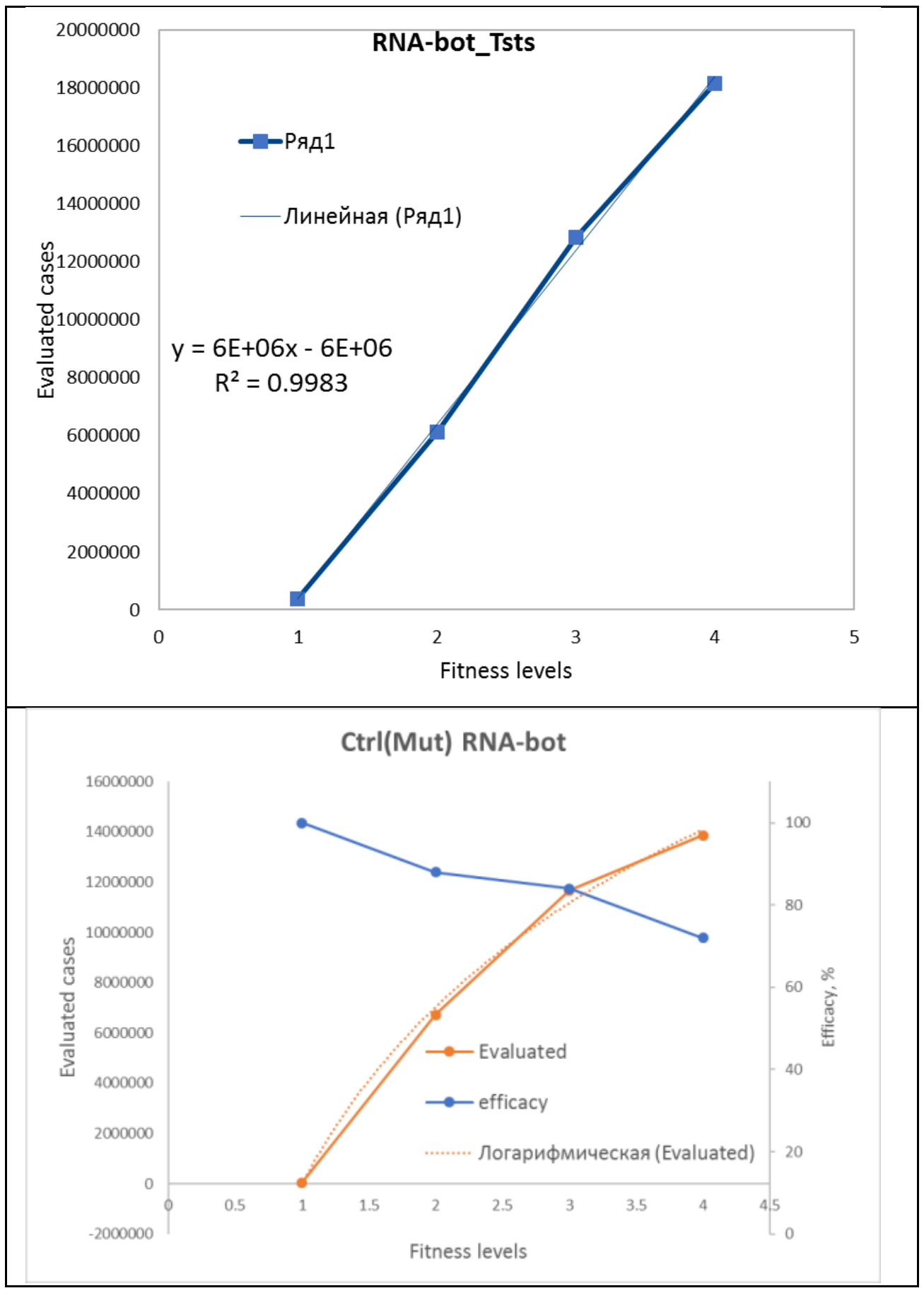

Figure 8. For an RNA device of 4 building blocks when searching from scratch, the speed of finding each next block by the retroGA algorithm is almost linear (A). Run Parameters: Popul=160000; HIVlim=8; RComb=27; _RpStp=20. (B) For tests with point mutations only, the same functional dependence is different from linear. Popul=480000; Quote=0.30; CrossGA=0.0100.

In this formulation, the problem is generally similar to the RS problem, while not RS, but RR under such conditions demonstrates the linearity of the search speed (as in fig.6).

We demonstrate that heuristic recombination algorithms are significantly more efficient in a test tube evolution model than point mutation. We believe that the implementation of new biochemical methods, based on such heuristic algorithms, can significantly improve the efficiency of *in vitro* evolution.

# 5 Discussion

## 5.1 Towards a theory of evolution of macromolecules modularity

A major aim with this work is to bridge evolutionary computations from benchmark cases, such as RR and RS, which are well-understood theoretically (in terms of mathematical analysis), to biological cases, which can serve as a basis for more efficient directed molecular evolution in the test tube and for understanding the mechanisms of biological evolution at the level of gene regulatory sequences.

Using analytical tools from statistical mechanics, dynamical systems theory, and mathematical population genetics, van Nimwegen and co-authors [van Nimwegen & Crutchfield, 2000; 2001; van Nimwegen et al., 1999; Crutchfield & Nimwegen, 2001] developed a detailed and quantitative description of the search dynamics for the RS and RR class of problems that exhibit epochal evolution.

More generally, the detailed understanding of the behavior for this class of problems provides valuable insights into the emergent mechanisms that control the dynamics in more general settings of evolutionary searches and in other population-based dynamical systems. By establishing the RR and RS characteristics of gene regulatory problems, we can use this theoretical background to anchor our understanding of more realistic biological search cases.

More recently, some other authors brought their attention to the RRF problem analysis. Brown with colleagues [Brown et al., 2002] have used the Markov Random Fields (MRFs) to construct an explicit probabilistic model of the RR problems. Then, Ter-Sarkissov and Marsland presented analysis of performance of an elitist Evolutionary algorithm using a recombination operator 1-Bit-Swap on the Royal Roads test function [Ter-Sarkissov and Marsland, 2011a; 2011b].

## 5.2 Implementation of heuristic algorithms in experimental procedures

The main methodological problem for the automated search for multimodal macromolecules from scratch in an experiment is selection procedures. High automation requires a selection system with an increasing number of ligands. Ligands, most likely, will need to be orderly fixed on the substrate.

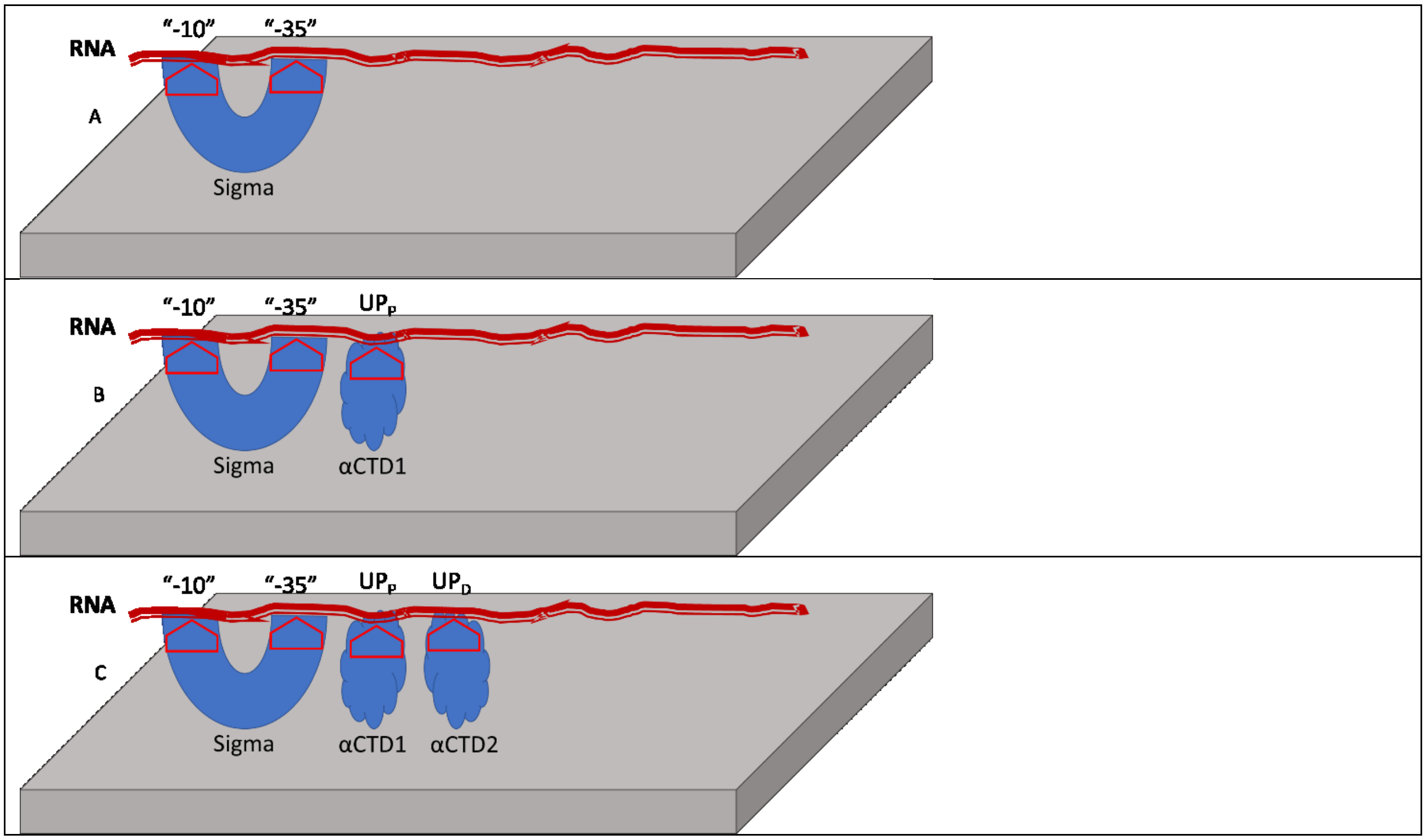

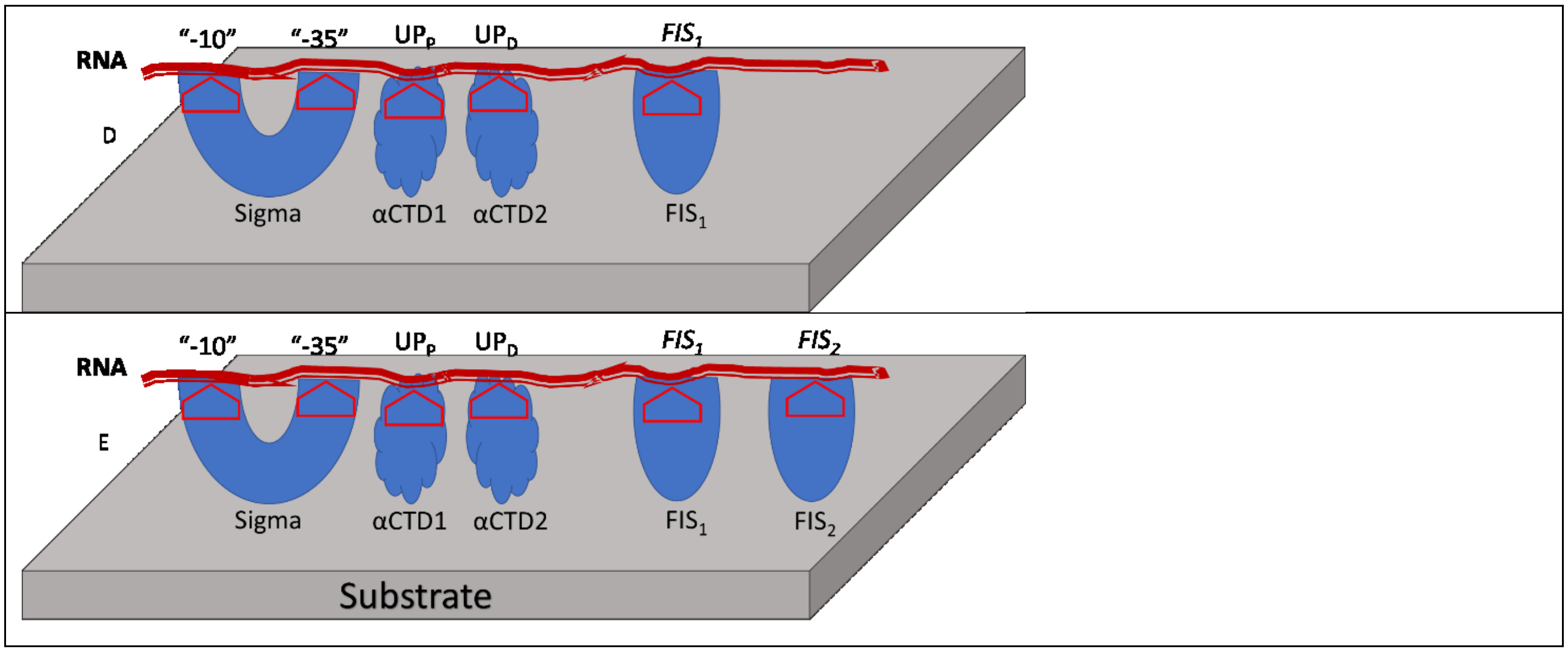

Figure 11. The general idea of step-by-step selection of motifs for a strong bacterial promoter. The procedure begins with selection for affinity to the (attached to the substrate) sigma subunit. Sequences with the desired level of affinity further, in the following rounds, are selected for affinity for a pair of adjacent sigma + alpha-CTD1 attached to the substrate. The found sequences with the desired affinity for both factors (with a triple of binding sites) are then selected for affinity for the next triple of factors located adjacent: sigma + alpha-CTD1 + alpha-CTD2. And so on.

A fairly simple search for motives using the example of a strong bacterial promoter is illustrated in Figure 11.

# 6 Conclusions

In conclusion, we want to draw the reader's attention to the fact that in this publication we tested the effectiveness of very old BB conservation algorithms, which were one of the first to be proposed. These were either procedures proposed by Mitchell et al. in the early-90s or procedures inspired by DNA shuffling and proposed by us a little later. As it turned out, the effectiveness of these procedures was at least half the order – one order of magnitude. Since then, a wide range of other algorithms have been developed that preserve BBs, both fairly general and highly specialized (for the traveling salesman problem, first of all). Therefore, it is natural to expect that next-generation algorithms, specially developed for many years for various optimization problems, may well be significantly more efficient (more than a dozen times more). Accordingly, their implementation in experimental in vitro evolutionary procedures can be very effective and significantly resource-saving.

# 7 Acknowledgments

The research was supported by the Russian Science Foundation (project No. 17-18-01536).